\documentclass[aps,twocolumn,showpacs,superscriptaddress]{revtex4}
\usepackage{multirow}
\usepackage{graphicx}
\usepackage{amssymb,amsmath}
\usepackage{bm} 
\usepackage{float}
\usepackage{xcolor}
\definecolor{midnightblue}{cmyk}{1,1,0,0.1}
\definecolor{forestation}{cmyk}{0.75,0,1,0.5}

\usepackage{hyperref}
\hypersetup{
    bookmarks=true,                 
    unicode=false,                     
    pdftoolbar=true,                   
    pdfmenubar=true,                
    pdffitwindow=false,              
    pdfstartview={Fitch},              
    pdftitle={My title},                  
    pdfauthor={Author},              
    pdfsubject={Subject},            
    pdfcreator={Creator},             
    pdfproducer={Producer},       
    pdfkeywords={keyword1} {key2} {key3}, 
    pdfnewwindow=true,             
    colorlinks=true,                      
    linkcolor=midnightblue,          
    citecolor=magenta,                
    filecolor=midnightblue,           
    urlcolor=midnightblue,            
}

\begin{document}
\title{Magnetic interactions in FeSe studied by first principle calculations}
\author{Shuai Wang}
\affiliation{International Center for Quantum Materials, School of Physics, Peking University, Beijing 100871, China}

\author{Fa Wang}
\affiliation{International Center for Quantum Materials, School of Physics, Peking University, Beijing 100871, China}
\affiliation{Collaborative Innovation Center of Quantum Matter, Beijing 100871, China}

\begin{abstract}
 Based on first principle calculations we have investigated the evolution of magnetism in  free-standing monolayer FeSe with respect to lattice constant and magnetism in bulk FeSe. The computational results show that the magnetic order in free-standing monolayer FeSe will change from anti-ferromagnetic pair-checkboard order to stripe collinear order along with enlarging lattice constant. The magnetic order in bulk FeSe will change from stripe collinear order to anti-ferromagnetic pair-checkboard order only if structure reconstruction is allowed.  We use J$_1$-J$_2$-K$_1$ model to fit the calculated total energies of different magnetic orders to study magnetic interaction strengths in FeSe. The fitting results of J$_1$-J$_2$-K$_1$ indicate that magnetic interactions in FeSe are quite strong and highly frustrated, and increase slowly with enlarging lattice parameter.
\end{abstract}
\maketitle
\section{Introduction}
In recent years up to 20meV superconducting gap  has been observed in monolayer FeSe grown on SrTiO$_3$(STO) substrate by scanning tunneling spectroscopy(STS)\cite{qing2012interface} and angle-resolved photoemission spectroscopy(ARPES)\cite{he2013phase,tan2013interface,liu2012electronic} experiments. Moreover, Jian-Feng Ge et al.\cite{ge2014superconductivity} reported the observation of over 100K superconductivity in FeSe/STO systems by in situ four-point probe electrical transport measurements, where the superconducting critical temperature(T$_c$) is almost twice of the T$_c$ record in FeAs-based supercondcutors\cite{zhi2008superconductivity,wang2008thorium} and ten times of T$_c$ measured in bulk FeSe under ambient pressure\cite{hsu2008superconductivity,PhysRevB.84.020503}.
Several possible mechanisms have been proposed to explain the T$_c$ enhancement.  The first possible mechanism  is charge transfer indicated by
the ARPES observation of only electron Fermi pockets in FeSe/STO\cite{he2013phase,tan2013interface,liu2012electronic}. The second one is strong electron-phonon coupling suggested by ARPES observation of "replica bands"\cite{lee2014interfacial}. In addition, stress caused by lattice mismatch between FeSe film and STO substrate is also thought to be one possible reason for T$_c$ enhancement\cite{PhysRevLett.112.107001}.
In bulk iron-based superconductors magnetic interactions seem to be important for superconductivity\cite{dai2012magnetism}. The parent compounds of iron pnictide superconductors have stripe collinear long range magnetic order, which has been confirmed by both experiments and Density Functional Theory(DFT) calculations\cite{de2008magnetic,zhao2008structural,0295-5075-83-2-27006,uemura2009superconductivity,ma2008iron,yildirim2008origin}. Although magnetic order has not been observed in FeSe, it is still possible that magnetism plays an important role in the T$_c$ enhancement in FeSe thin film. Previous studies have investigated the effect of stress on magnetism in FeSe, however the reported magnetic interaction strengths are inconsistent\cite{cao2014interfacial,glasbrenner2015effect,PhysRevB.91.020504}. Therefore our main motivation is to reexamine the influence of stress on electronic structure and magnetism of monolayer FeSe.

In this paper, we have studied the evolution of magnetism in free-standing monolayer FeSe with respect to lattice constant and magnetism in bulk FeSe\cite{PhysRevLett.103.057002} by means of first principle calculations. Lattice constants 3.765{\AA}, 3.905{\AA} and 4.045{\AA} were chosen for monolayer FeSe during the calculations. The three kinds of lattice constants were taken from previous DFT work\cite{cao2014interfacial}, which are supposed to be close to lattice constants of bulk FeSe, STO substrate and KTaO$_3$ substrate respectively.  The calculated results show that anti-ferromagnetic pair-checkboard order is magnetic ground state of free-standing monolayer FeSe with lattice constant 3.765{\AA} and bulk FeSe under structure reconstruction, and that stripe collinear order is magnetic ground state of free-standing monolayer FeSe with lattice constant 3.905{\AA} and 4.045{\AA}. The anti-ferromagnetic pair-checkboard order in bulk FeSe was first found by Hai-Yuan Cao et al.\cite{PhysRevB.91.020504} via first principle calculations. Furthermore, we used $\text{J}_1$-$\text{J}_2$-$\text{K}_1$ model\cite{YareskoInterplay,wysocki2011consistent}  to fit calculated total energies of different magnetic orders to study magnetic interactions in FeSe and the relations between magnetic interaction and lattice constant. The results indicate that magnetic interaction strengths increase slowly as enlarging lattice constant. The ratio of fitted $\text{J}_2$/$\text{J}_1$ is close to 0.5 for both bulk FeSe and monolayer FeSe, which implies strong magnetic frustration in FeSe. Based on J$_1$-J$_2$-K$_1$ model with fitted exchange parameters we calculated spin wave dispersions of bulk FeSe, which can be tested in inelastic neutron scattering experiments.

\section{Methods}
\begin{figure}
\begin{centering}
\includegraphics[width=0.49\textwidth]{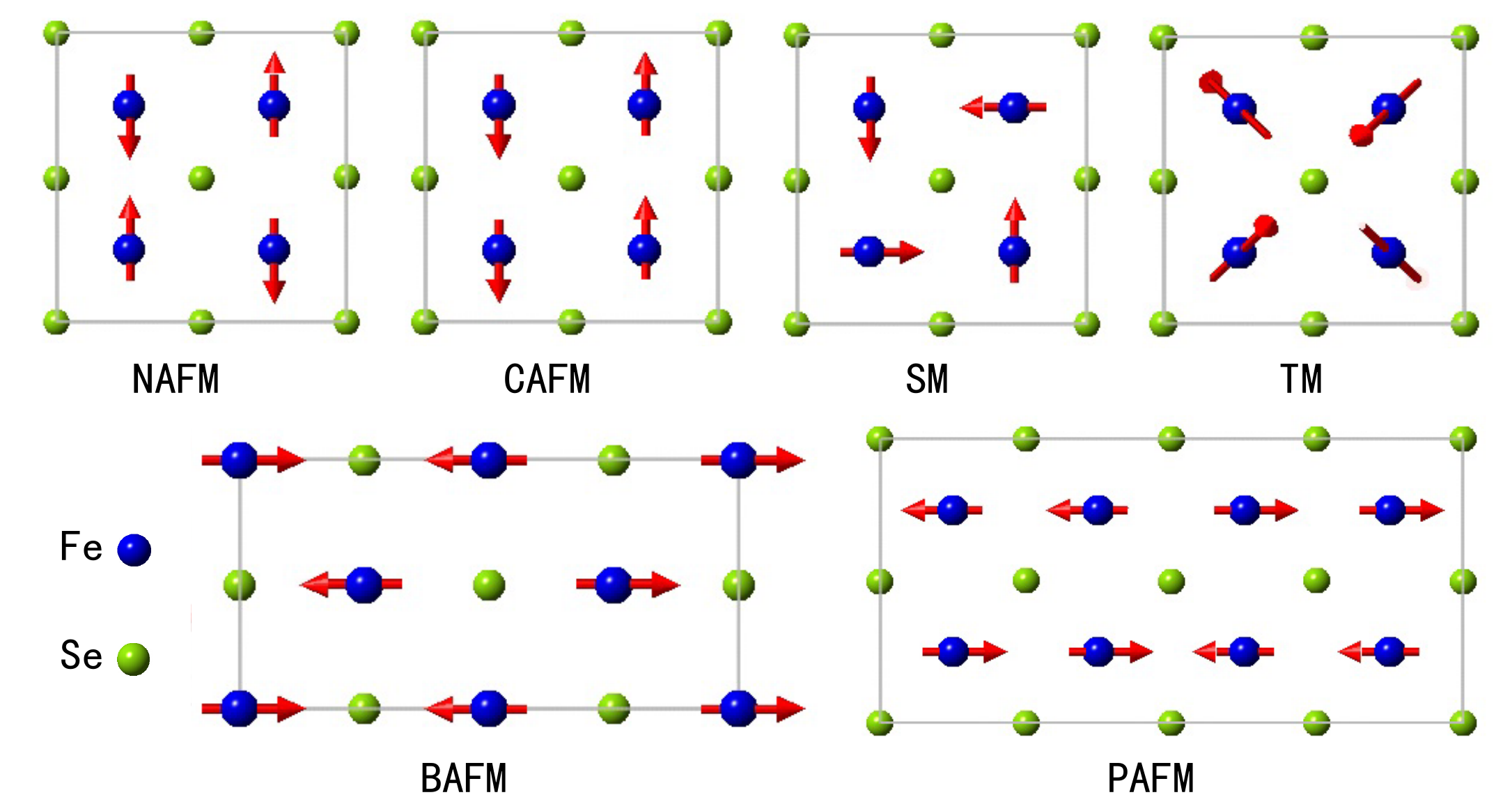}
\end{centering}
\caption{Schematic diagram of different magnetic orders in top view. NAFM, CAFM, SM, TM, BAFM, PAFM represent N\'eel anti-ferromagnetic order, collinear anti-ferromagnetic order, spiral magnetic order, tetrahedron magnetic order, bi-collinear anti-ferromagnetic order and pair-checkboard anti-ferromagnetic order respectively. The tetrahedron magnetic order means that the moments of the four Fe atoms in the unit cell form an tetrahedron. The solid grey lines enclose magnetic unit cell of corresponding magnetic order used in the calculations.}\label{magneticorder}
\end{figure}

Non-collinear magnetic calculations implemented in Vienna Ab initio Simulation Package(VASP)\cite{PhysRevB.62.11556,Kresse199615,PhysRevB.54.11169} were performed to study the electronic and magnetic structures of free-standing monolayer FeSe with different lattice constants and bulk FeSe. The project augmented wave(PAW) pseudopotential\cite{PhysRevB.50.17953,PhysRevB.59.1758}with generalized gradient approximation(GGA) of Perdew-Burke-Ernzerhof(PBE)\cite{PhysRevLett.77.3865} for exchange-correlation potential were adopted in the calculations. We used 24$\times$24$\times$1 \textbf{k}-mesh and 18$\times 18\times 18$ \textbf{k}-mesh for reciprocal space sampling in monolayer FeSe and bulk FeSe respectively\cite{PhysRevB.13.5188}, 500eV for plane wave energy cut-off, and Gaussian Smearing with broadening width of 0.01eV to capture converged results in meV scale. Over 15{\AA} vacuum layer was added to decouple interlayer couplings. We chose $\sqrt{2}\times\sqrt{2}\times1$ supercell in the calculations of non-magnetic(NM) state, N\'eel anti-ferromagnetic(NAFM) state, collinear anti-ferromagnetic(CAFM) state, spiral magnetic(SM) state, and tetrahedron magnetic(TM) state. $2\times1\times1$ supercell and $2\sqrt{2}\times\sqrt{2}\times1$ supercell  were chosen in the calculations of bi-collinear anti-ferromagnetic(BAFM) state and anti-ferromagnetic pair-checkboard(PAFM) state respectively. All the calculated magnetic states are schematically shown in the Figure \ref{magneticorder}. We used the following frustrated J$_1$-J$_2$-K$_1$\cite{YareskoInterplay,wysocki2011consistent}  model to describe magnetic interactions, where the superexchange parameters are fitted to the total energies with variable moment sizes of different magnetic orders.
\begin{equation}\label{J1J2K1formula}
\begin{aligned}
H_{spin} =&\sum_{\langle i,j\rangle}{J_1 \textbf{S}_i\cdot\textbf{S}_j}+\sum_{\langle\langle i,j\rangle\rangle}{J_2\textbf{S}_i\cdot\textbf{S}_j}-\sum_{\langle i,j\rangle}{K_1[\textbf{S}_i\cdot \textbf{S}_j]^2}\\
  =&\sum_{\langle i,j\rangle}{\frac{J_1}{g^2\mu^2_{B}} \textbf{M}_i\cdot\textbf{M}_j}+\sum_{\langle\langle i,j\rangle\rangle}{\frac{J_2}{g^2\mu^2_{B}}\textbf{M}_i\cdot\textbf{M}_j}\\
  -&\sum_{\langle i,j\rangle}{\frac{K_1}{g^4\mu^4_{B}}[\textbf{M}_i\cdot \textbf{M}_j]^2}
\end{aligned}
\end{equation}
where $\textbf{M}_i$ ($\textbf{M}_j$) labels the magnetic moment of atom on site $i$ ($j$). $\langle i,j\rangle, \langle\langle i,j\rangle\rangle$ denote all the summations over nearest, next-nearest neighbor respectively. $g$ labels the Land\'{e} g-factor. $\mu_{B}$ represents Bohr magneton.

\section{Results and Discussion}
\begin{figure}
  \includegraphics[width=0.49\textwidth]{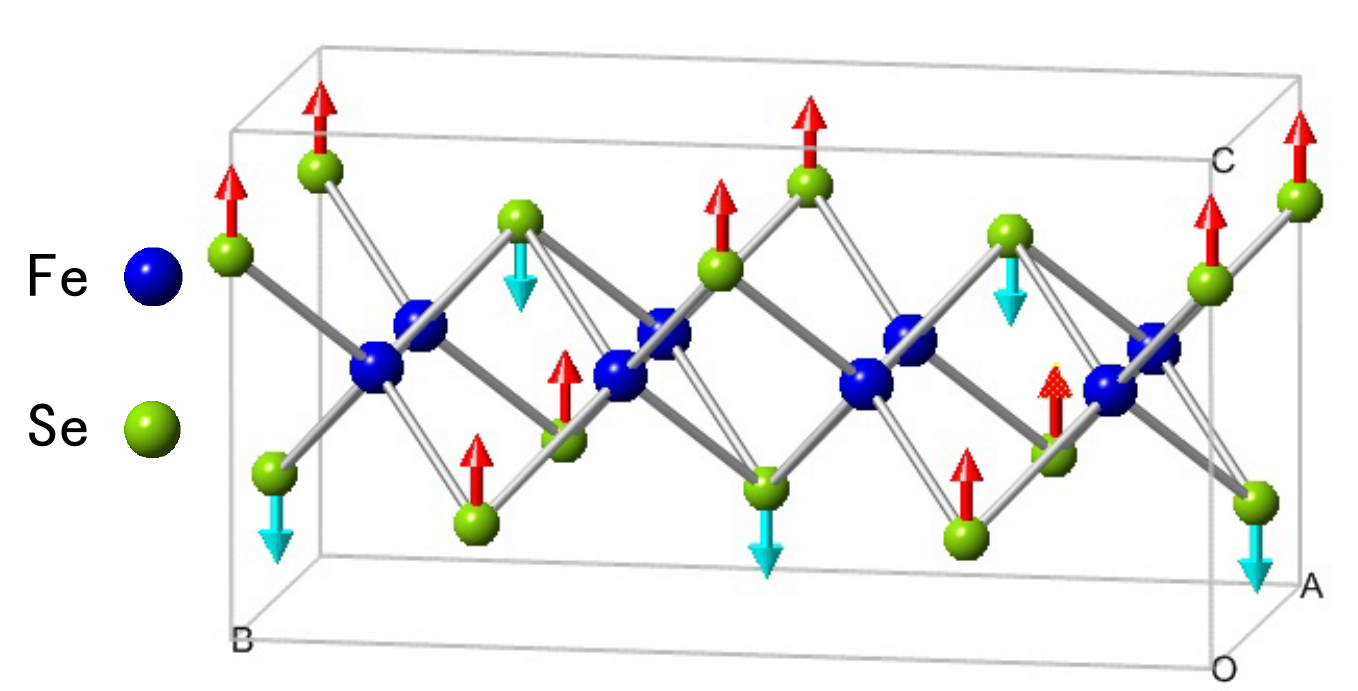}\\
  \caption{Se atoms displacement related to the structure reconstruction in FeSe under PAFM order. The solid grey lines enclose magnetic unit cell of FeSe with PAFM order.}\label{Sedisplacement}
\end{figure}

In the calculations we used crystal structure after only relaxing ionic positions under CAFM order. Since the energy difference in bulk FeSe is quite small(1.5 meV per Fe atom under CAFM order) between ferromagnetic stacking and anti-ferromagnetic stacking in the c axis direction, we just considered ferromagnetic stacking in the c axis direction for simplicity during the calculations.  All the calculated total energies and moments of free-standing monolayer FeSe with different lattice constants and bulk FeSe are illustrated in the Table \ref{totalenergy}.
\begin{table}[H]
\centering
\begin{tabular}{|c|c|c|c|c|c|}
  \hline
\multicolumn{2}{|c|}{a({\AA})} & 3.765 & 3.905 & 4.045 &bulk \\
\hline
\multicolumn{2}{|c|}{$Z_{Se}$({\AA})}  & 1.444&1.393&1.326&1.418\\
\hline
\multirow{2}{*}{NAFM}&E(eV)&-49.628&-49.374&-48.968 & -49.547 \\
\cline{2-6}
   & M($\mu_{B}$) & 1.94& 2.22 & 2.42 & 1.88\\
\hline
\multirow{2}{*}{CAFM} & E(eV) & -49.761& -49.647 & -49.396 &-49.690 \\
\cline{2-6}
   & M($\mu_{B}$)& 2.13& 2.42 & 2.63 & 2.05 \\
\hline
\multirow{2}{*}{PAFM} & E(eV)& -49.808& -49.614& -49.234 & $-49.728^\star$ \\
\cline{2-6}
   & M($\mu_{B}$) & 2.12 & 2.35& 2.52& $2.09^\star$ \\
\hline
\multirow{2}{*}{BAFM}& E(eV) & -49.392& -49.122& -48.736 & -49.333\\
\cline{2-6}
   & M($\mu_{B}$) & 2.01 & 2.35 & 2.67 & 1.89\\
\hline
\multirow{2}{*}{SM}& E(eV) & -49.505 & -49.366 & -49.143& -49.458\\
\cline{2-6}
   & M($\mu_{B}$) & 1.96 & 2.43 & 2.63 & 1.80 \\
\hline
\multirow{2}{*}{TM} & E(eV) & -49.548 & -49.346 & -49.033& -49.394 \\
\cline{2-6}
   & M($\mu_{B}$) & 1.89 & 2.29 & 2.52 & 1.90 \\
  \hline
\end{tabular}\caption{Total energies and magnetic moments for free-standing monolayer FeSe with different lattice constants and bulk FeSe under different magnetic orders. The total energies are given with respect to  magnetic unit cell with 4 Fe atoms. For non-collinear TM order and SM order, the directions of converged moments have tiny difference from the originally set directions(up to 2 degree). Total energies of NM states for monolayer FeSe with lattice constants 3.765{\AA}, 3.905{\AA}, 4.045{\AA}, and bulk FeSe are -49.113eV, -48.638eV, -47.960eV, -49.146eV respectively. Energy and moment of  bulk FeSe  labeled with $\star$  are obtained after relaxing the height of Se atoms. Due to the reconstruction under PAFM order, the total energies and moments under PAFM order were not used in fitting exchange parameters.  }\label{totalenergy}
\end{table}
As the results shown in the Table \ref{totalenergy}, PAFM state  is stable in monolayer FeSe with lattice constant 3.765{\AA}, where the energy is lower than CAFM order by 12meV per Fe atom. The total energies of CAFM order are still lowest among calculated states in monolayer FeSe with lattice constant 3.905{\AA}, 4.045{\AA}, and in bulk FeSe. The magnetic state with initial PAFM order in bulk FeSe converges to NM state after static calculations performed in VASP. When optimizing ionic positions under PAFM order, we found that  there is a $2\times1$ structure reconstruction in the long axis of magnetic unit cell for both monolayer FeSe with the three kinds of lattice constants and bulk FeSe.
The reconstruction comes from the different distances of Se atoms to Fe atoms plane, which has been schematically shown in the Figure \ref{Sedisplacement}. In bulk FeSe the reconstruction leads to the magnetic ground state changing from CAFM order to PAFM order with total energy lower by 9.5meV per Fe.

Although no long range magnetic order has been observed in bulk FeSe, it does not  mean that the magnetic interaction can been ignored. Previous studies in iron-based superconductors used local spin model to describe the magnetic interactions\cite{yildirim2008origin,si2008strong,yaomagnetic,fangtheory,ma2008arsenic}. Here we also employ  a frustrated J$_1$-J$_2$-K$_1$ model\cite{YareskoInterplay,wysocki2011consistent} to describe the magnetic interactions. Since the values of moment sizes are not strictly 2 $\mu_{B}$, we fitted total energies with variable moment sizes to obtain superexhange parameters (see equation (\ref{J1J2K1formula})). Table \ref{J1J2K1} lists the fitting results.
\begin{table}[H]
\centering
\begin{tabular}{|c|c|c|c|c|}
  \hline
a({\AA})& 3.765 & 3.905& 4.045& bulk\\
\hline
$Z_{Se}$({\AA})& 1.444& 1.393 & 1.326&1.418\\
\hline
$J_1/g^2(meV)$  &21.5$\pm$2.8 & 24.0$\pm$3.2&  29.5$\pm$4.9&17.9$\pm$1.7\\
\hline
$J_2/g^2(meV)$  & 10.9$\pm$1.4 & 13.2$\pm$1.6& 17.4$\pm$2.4&9.9$\pm$0.9 \\
\hline
$K_1/g^4(meV)$   &  1.4$\pm$0.3 &1.3$\pm$0.3 & 1.2$\pm$0.3&1.5$\pm$0.2\\
\hline
$J_2/J_1$       & 0.51 & 0.55&  0.59 &0.55\\
\hline
\end{tabular}\caption{Fitting values of J$_1$, J$_2$, K$_1$ for monolayer FeSe with different lattice constants and bulk FeSe. }\label{J1J2K1}
\end{table}
The fitting values of superexchange parameters have been divided by the square or fourth power of  Land\'{e} g-factor. If the Land\'{e} g-factor is 2, the given exchange parameters should multiple by 4 or 16. The data in the Table \ref{J1J2K1} indicates that the magnetic interactions in FeSe are quite strong and that magnetic interactions in monolayer FeSe is stronger than that in bulk FeSe. We can also see that the  magnetic interaction strengths increase slowly with enlarging lattice parameter in monolayer FeSe.

What's more, the fitting ratio of $\text{J}_2$/$\text{J}_1$ in both monolayer FeSe and bulk FeSe is close to 0.5. It is known in $\text{J}_1$-$\text{J}_2$ heisenberg model of square lattice that the system becomes highly frustrated as the ratio of J$_2$/J$_1$ close to 0.5. Magnetic frustration has been proposed as the explanation for the absence of magnetic long range order in bulk FeSe\cite{PhysRevB.79.174409,wang2015nematicity,glasbrenner2015effect,yuantiferroquadrupolar,PhysRevB.91.201105}.
So although magnetic order can be obtained in FeSe with DFT calculations, the strong quantum fluctuation can destroy the long range magnetic order\cite{glasbrenner2015effect}.  J. K. Glasbrenner et al.\cite{glasbrenner2015effect} and Zhong-Yi Lu et al.\cite{Xproceedings} found that there exist a series of magnetic states between PAFM order and CAFM order, which also imply strong magnetic frustration in FeSe.

\begin{figure}[H]
  \includegraphics[scale=0.5]{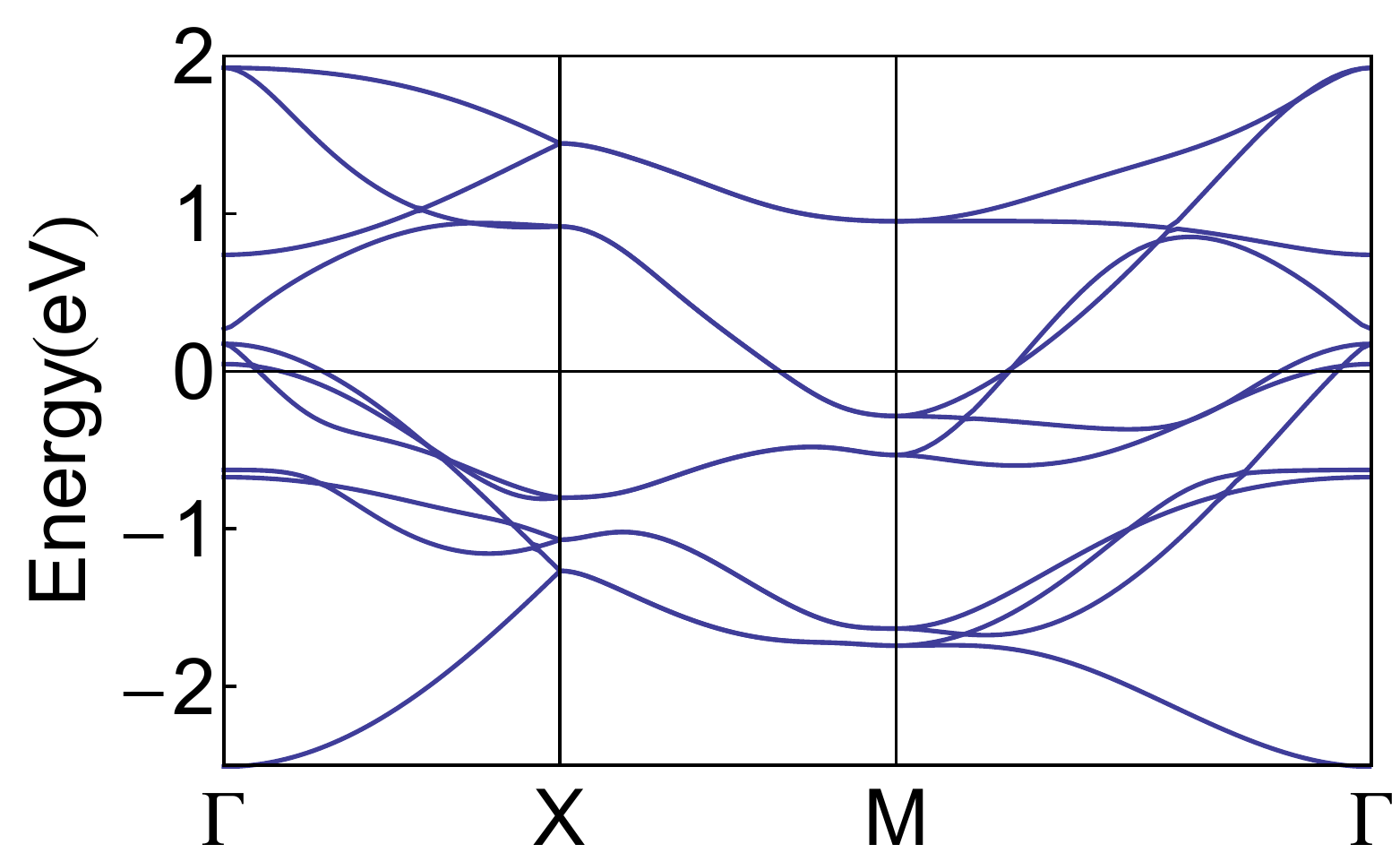}
   \includegraphics[scale=0.5]{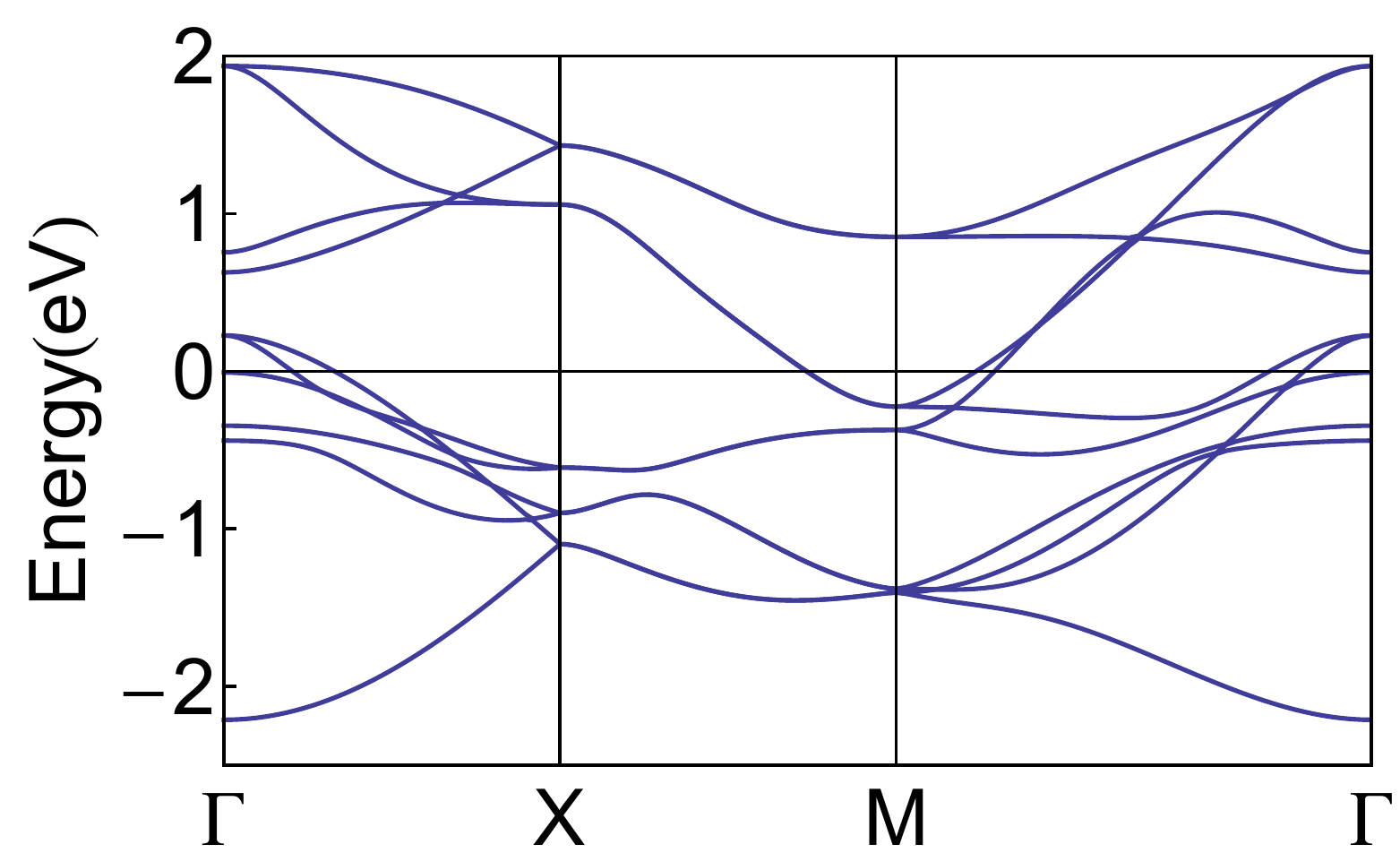}
    \includegraphics[scale=0.5]{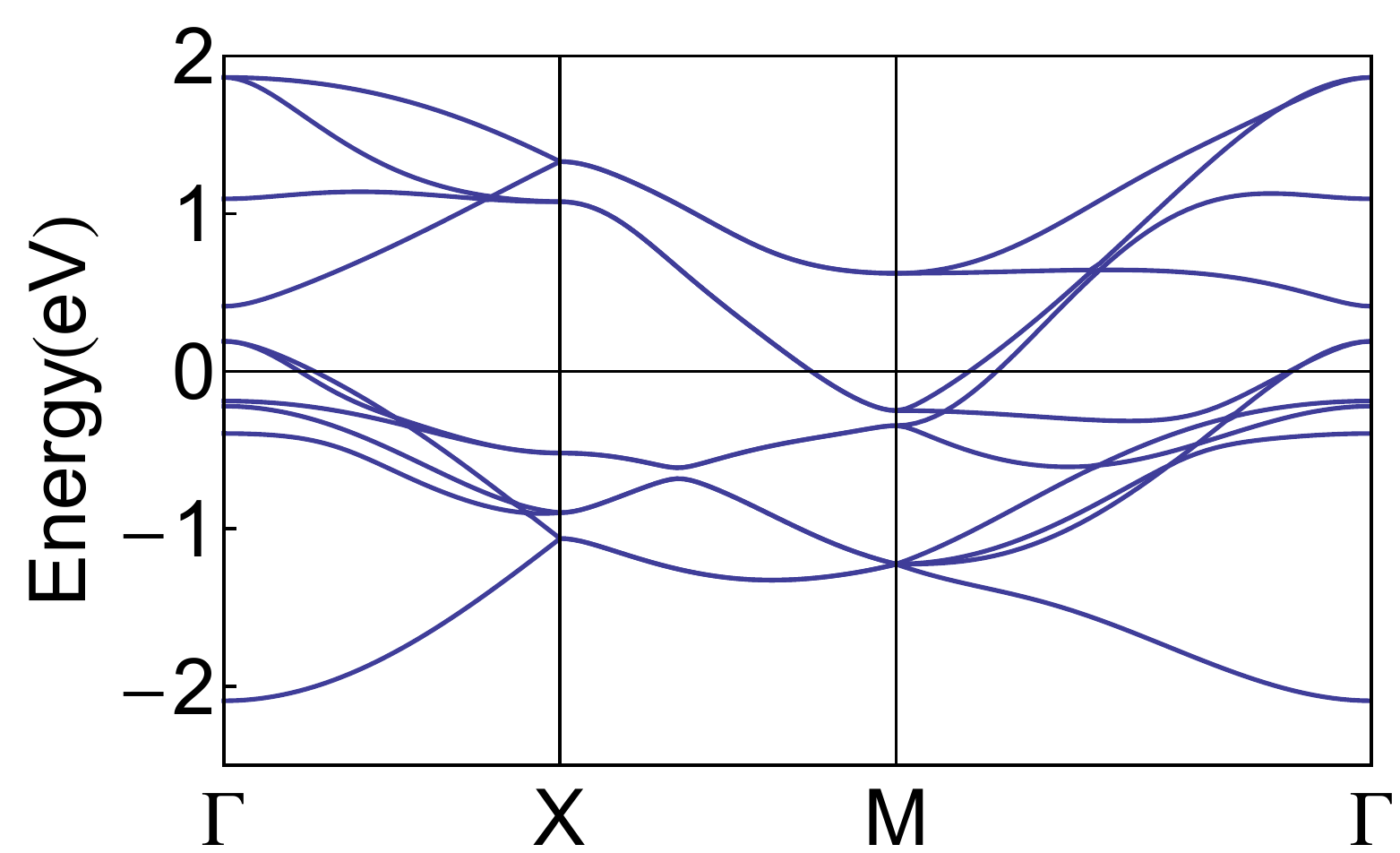}\\
  \caption{Band structures of free-standing monolayer FeSe with lattice constant a= 3.765{\AA}, 3.905{\AA} and 4.045{\AA}. The band structures are calculated under NM states. The coordinates of  special points $\Gamma$, X, M are (0,0), ($\pi$/a,0), and ($\pi$/a,$\pi$/a) respectively.}\label{bandstructure}
\end{figure}
We also calculated electronic structures of monolayer FeSe with the three kinds of lattice constants under NM state to study lattice constant's influence on band structure. The band structures shown in the Figure \ref{bandstructure} indicate that enlarging lattice parameter does not change band structures significantly, except a small change in bandwidth and a small hole fermi pocket located at $\Gamma$ point dropping below fermi level. The fact that band structures and magnetic interactions do not dramatically change with enlarging lattice constant implies that T$_c$ enhancement in FeSe/STO is unlikely induced by tensile stress.

To test the validity and accuracy of our approach to fitting exchange parameters, we calculated spin wave dispersions in bulk FeSe based on linear spin wave theory, which can be experimentally measured by inelastic neutron scattering. Since no structure reconstruction in bulk FeSe has been observed by experiments, we then chose CAFM order as magnetic ground state during the spin wave calculations. The magnon dispersions within J$_1$-J$_2$-K$_1$ model can be written as,
\begin{equation}\label{spinwave}
E_{LSW}(\textbf{k})=\sqrt{A_\textbf{k}^2-B_\textbf{k}^2},
\end{equation}
where
\begin{equation}
\begin{aligned}
  A_{\textbf{k}} &= (2J_1cos(k_y)+4J_2)S+2K_1(4-2cos(k_y))S^3,\\
  B_{\textbf{k}} &= (2J_1cos(k_x)+4J_2cos(k_x)cos(k_y))S+4K_1cos(k_x)S^3.
\end{aligned}
\end{equation}
\begin{figure}[H]
  \includegraphics[width=0.49\textwidth]{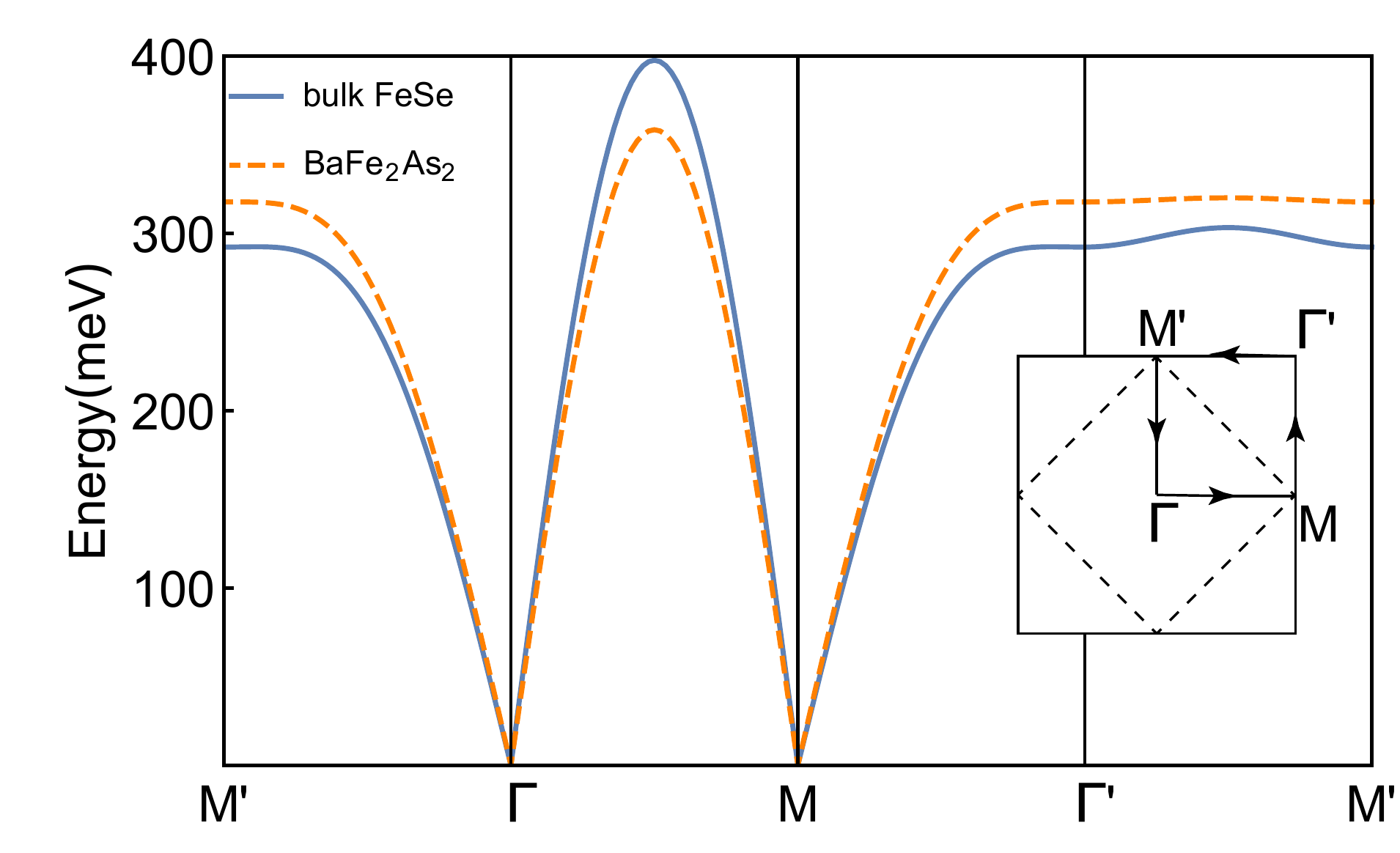}\\
  \caption{Spin wave dispersions of tetragonal BaFe$_2$As$_2$ and bulk FeSe under stripe collinear order, where we chose g=2 and S=1 . The dispersions are drawn in the unfolded Brillouin zone(one Fe atom per unit cell), where the high symmetry points are shown in the inset. }\label{spinwave}
\end{figure}
For comparison we also calculated spin wave dispersions of tetragonal BaFe$_2$As$_2$\cite{huang2008neutron} within J$_1$-J$_2$-K$_1$ model, where J$_1$ and J$_2$ are also obtained by fitting total energies with variable moment sizes. The fitted values are J$_1$/g$^2$=13.6$\pm$2.4meV,  J$_2$/g$^2$=10.0$\pm$1.1meV and K$_1$/g$^4=$1.4$\pm$0.3meV. The ratio of J$_2$/J$_1$ is around 0.74, which is larger than that in both monolayer FeSe and bulk FeSe, and thus make CAFM order stable. The magnon dispersions along high symmetry points are shown in the Figure \ref{spinwave}. We can see that bulk FeSe has a slightly larger bandwidth of magnon dispersions compared with BaFe$_2$As$_2$, assuming the same $g$-factor and spin size for the two compunds. Although no magnetic long range order has been observed in bulk FeSe, we hope that the spin wave dispersions under CAFM order can still qualitatively describe the magnetic excitations in bulk FeSe.

We noticed that our fitting results for monolayer FeSe are different from previous calculations\cite{cao2014interfacial}, where the following reasons may be responsible for that. (\textbf{i}) We used more dense k-mesh and smaller broadening width compared with their calculations. (\textbf{ii}) Non-collinear magnetic states(SM and TM) were included in our calculations. (\textbf{iii}) We calculated the total energies and moments of different magnetic orders with fixed structure to fit the exchange parameters J$_1$, J$_2$, K$_1$, where they optimized structure under different magnetic orders during the calculations. (\textbf{iv}) In addition, here we have used the method considering the magnetic moment sizes to fit exchange parameters, which were not taken into account in their fittings.
Our J$_1$-J$_2$-K$_1$ fitting results for bulk FeSe are in rough agreement with the results of F. Ma et al.\cite{PhysRevLett.102.177003}, but are significantly smaller than the results of J. Glasbrenner et al.\cite{glasbrenner2015effect}.

In conclusion, the computational results show that the magnetic order in monolayer FeSe will change from PAFM order to CAFM order with enlarging lattice constant, and that the magnetic order in bulk FeSe will change from CAFM order to PAFM order only if structure reconstruction is  allowed. The fitted exchange parameters within J$_1$-J$_2$-K$_1$ model suggest strong frustration in monolayer FeSe and bulk FeSe, which may provide hint to understand why long range magnetic order has not been observed in FeSe. According to the results, the bandwidth of spin wave dispersions in bulk FeSe would be slightly larger than that in BaFe$_2$As$_2$, which can be verified in inelastic neutron scattering experiments.

\textbf{Acknowledgements:}
The computational work was performed on TianHe-1(A) at National Superconductor Center in Tianjin. Shuai Wang thanks Long Zhang for helpful discussions.  This work was supported by the National Key Basic Research Program of China (Grant No. 2014CB920902) and the National Science Foundation of China (Grant No. 11374018).
\bibliography{refs}

\begin{thebibliography}{42}
\expandafter\ifx\csname natexlab\endcsname\relax\def\natexlab#1{#1}\fi
\expandafter\ifx\csname bibnamefont\endcsname\relax
  \def\bibnamefont#1{#1}\fi
\expandafter\ifx\csname bibfnamefont\endcsname\relax
  \def\bibfnamefont#1{#1}\fi
\expandafter\ifx\csname citenamefont\endcsname\relax
  \def\citenamefont#1{#1}\fi
\expandafter\ifx\csname url\endcsname\relax
  \def\url#1{\texttt{#1}}\fi
\expandafter\ifx\csname urlprefix\endcsname\relax\def\urlprefix{URL }\fi
\providecommand{\bibinfo}[2]{#2}
\providecommand{\eprint}[2][]{\url{#2}}

\bibitem[{\citenamefont{Qing-Yan et~al.}(2012)\citenamefont{Qing-Yan, Zhi,
  Wen-Hao, Zuo-Cheng, Jin-Song, Wei, Hao, Yun-Bo, Peng, Kai
  et~al.}}]{qing2012interface}
\bibinfo{author}{\bibfnamefont{W.}~\bibnamefont{Qing-Yan}},
  \bibinfo{author}{\bibfnamefont{L.}~\bibnamefont{Zhi}},
  \bibinfo{author}{\bibfnamefont{Z.}~\bibnamefont{Wen-Hao}},
  \bibinfo{author}{\bibfnamefont{Z.}~\bibnamefont{Zuo-Cheng}},
  \bibinfo{author}{\bibfnamefont{Z.}~\bibnamefont{Jin-Song}},
  \bibinfo{author}{\bibfnamefont{L.}~\bibnamefont{Wei}},
  \bibinfo{author}{\bibfnamefont{D.}~\bibnamefont{Hao}},
  \bibinfo{author}{\bibfnamefont{O.}~\bibnamefont{Yun-Bo}},
  \bibinfo{author}{\bibfnamefont{D.}~\bibnamefont{Peng}},
  \bibinfo{author}{\bibfnamefont{C.}~\bibnamefont{Kai}}, \bibnamefont{et~al.},
  \bibinfo{journal}{Chinese Physics Letters} \textbf{\bibinfo{volume}{29}},
  \bibinfo{pages}{037402} (\bibinfo{year}{2012}).

\bibitem[{\citenamefont{He et~al.}(2013)\citenamefont{He, He, Zhang, Zhao, Liu,
  Liu, Mou, Ou, Wang, Li et~al.}}]{he2013phase}
\bibinfo{author}{\bibfnamefont{S.}~\bibnamefont{He}},
  \bibinfo{author}{\bibfnamefont{J.}~\bibnamefont{He}},
  \bibinfo{author}{\bibfnamefont{W.}~\bibnamefont{Zhang}},
  \bibinfo{author}{\bibfnamefont{L.}~\bibnamefont{Zhao}},
  \bibinfo{author}{\bibfnamefont{D.}~\bibnamefont{Liu}},
  \bibinfo{author}{\bibfnamefont{X.}~\bibnamefont{Liu}},
  \bibinfo{author}{\bibfnamefont{D.}~\bibnamefont{Mou}},
  \bibinfo{author}{\bibfnamefont{Y.-B.} \bibnamefont{Ou}},
  \bibinfo{author}{\bibfnamefont{Q.-Y.} \bibnamefont{Wang}},
  \bibinfo{author}{\bibfnamefont{Z.}~\bibnamefont{Li}}, \bibnamefont{et~al.},
  \bibinfo{journal}{Nature materials} \textbf{\bibinfo{volume}{12}},
  \bibinfo{pages}{605} (\bibinfo{year}{2013}).

\bibitem[{\citenamefont{Tan et~al.}(2013)\citenamefont{Tan, Zhang, Xia, Ye,
  Chen, Xie, Peng, Xu, Fan, Xu et~al.}}]{tan2013interface}
\bibinfo{author}{\bibfnamefont{S.}~\bibnamefont{Tan}},
  \bibinfo{author}{\bibfnamefont{Y.}~\bibnamefont{Zhang}},
  \bibinfo{author}{\bibfnamefont{M.}~\bibnamefont{Xia}},
  \bibinfo{author}{\bibfnamefont{Z.}~\bibnamefont{Ye}},
  \bibinfo{author}{\bibfnamefont{F.}~\bibnamefont{Chen}},
  \bibinfo{author}{\bibfnamefont{X.}~\bibnamefont{Xie}},
  \bibinfo{author}{\bibfnamefont{R.}~\bibnamefont{Peng}},
  \bibinfo{author}{\bibfnamefont{D.}~\bibnamefont{Xu}},
  \bibinfo{author}{\bibfnamefont{Q.}~\bibnamefont{Fan}},
  \bibinfo{author}{\bibfnamefont{H.}~\bibnamefont{Xu}}, \bibnamefont{et~al.},
  \bibinfo{journal}{Nature materials} \textbf{\bibinfo{volume}{12}},
  \bibinfo{pages}{634} (\bibinfo{year}{2013}).

\bibitem[{\citenamefont{Liu et~al.}(2012)\citenamefont{Liu, Zhang, Mou, He, Ou,
  Wang, Li, Wang, Zhao, He et~al.}}]{liu2012electronic}
\bibinfo{author}{\bibfnamefont{D.}~\bibnamefont{Liu}},
  \bibinfo{author}{\bibfnamefont{W.}~\bibnamefont{Zhang}},
  \bibinfo{author}{\bibfnamefont{D.}~\bibnamefont{Mou}},
  \bibinfo{author}{\bibfnamefont{J.}~\bibnamefont{He}},
  \bibinfo{author}{\bibfnamefont{Y.-B.} \bibnamefont{Ou}},
  \bibinfo{author}{\bibfnamefont{Q.-Y.} \bibnamefont{Wang}},
  \bibinfo{author}{\bibfnamefont{Z.}~\bibnamefont{Li}},
  \bibinfo{author}{\bibfnamefont{L.}~\bibnamefont{Wang}},
  \bibinfo{author}{\bibfnamefont{L.}~\bibnamefont{Zhao}},
  \bibinfo{author}{\bibfnamefont{S.}~\bibnamefont{He}}, \bibnamefont{et~al.},
  \bibinfo{journal}{Nature communications} \textbf{\bibinfo{volume}{3}},
  \bibinfo{pages}{931} (\bibinfo{year}{2012}).

\bibitem[{\citenamefont{Ge et~al.}(2014)\citenamefont{Ge, Liu, Liu, Gao, Qian,
  Xue, Liu, and Jia}}]{ge2014superconductivity}
\bibinfo{author}{\bibfnamefont{J.-F.} \bibnamefont{Ge}},
  \bibinfo{author}{\bibfnamefont{Z.-L.} \bibnamefont{Liu}},
  \bibinfo{author}{\bibfnamefont{C.}~\bibnamefont{Liu}},
  \bibinfo{author}{\bibfnamefont{C.-L.} \bibnamefont{Gao}},
  \bibinfo{author}{\bibfnamefont{D.}~\bibnamefont{Qian}},
  \bibinfo{author}{\bibfnamefont{Q.-K.} \bibnamefont{Xue}},
  \bibinfo{author}{\bibfnamefont{Y.}~\bibnamefont{Liu}}, \bibnamefont{and}
  \bibinfo{author}{\bibfnamefont{J.-F.} \bibnamefont{Jia}},
  \bibinfo{journal}{Nature materials} \textbf{\bibinfo{volume}{14}},
  \bibinfo{pages}{285} (\bibinfo{year}{2014}).

\bibitem[{\citenamefont{Zhi-An et~al.}(2008)\citenamefont{Zhi-An, Wei, Jie,
  Wei, Xiao-Li, Guang-Can, Xiao-Li, Li-Ling, Fang, Zhong-Xian
  et~al.}}]{zhi2008superconductivity}
\bibinfo{author}{\bibfnamefont{R.}~\bibnamefont{Zhi-An}},
  \bibinfo{author}{\bibfnamefont{L.}~\bibnamefont{Wei}},
  \bibinfo{author}{\bibfnamefont{Y.}~\bibnamefont{Jie}},
  \bibinfo{author}{\bibfnamefont{Y.}~\bibnamefont{Wei}},
  \bibinfo{author}{\bibfnamefont{S.}~\bibnamefont{Xiao-Li}},
  \bibinfo{author}{\bibfnamefont{C.}~\bibnamefont{Guang-Can}},
  \bibinfo{author}{\bibfnamefont{D.}~\bibnamefont{Xiao-Li}},
  \bibinfo{author}{\bibfnamefont{S.}~\bibnamefont{Li-Ling}},
  \bibinfo{author}{\bibfnamefont{Z.}~\bibnamefont{Fang}},
  \bibinfo{author}{\bibfnamefont{Z.}~\bibnamefont{Zhong-Xian}},
  \bibnamefont{et~al.}, \bibinfo{journal}{Chinese Physics Letters}
  \textbf{\bibinfo{volume}{25}}, \bibinfo{pages}{2215} (\bibinfo{year}{2008}).

\bibitem[{\citenamefont{Wang et~al.}(2008)\citenamefont{Wang, Li, Chi, Zhu,
  Ren, Li, Wang, Lin, Luo, Jiang et~al.}}]{wang2008thorium}
\bibinfo{author}{\bibfnamefont{C.}~\bibnamefont{Wang}},
  \bibinfo{author}{\bibfnamefont{L.}~\bibnamefont{Li}},
  \bibinfo{author}{\bibfnamefont{S.}~\bibnamefont{Chi}},
  \bibinfo{author}{\bibfnamefont{Z.}~\bibnamefont{Zhu}},
  \bibinfo{author}{\bibfnamefont{Z.}~\bibnamefont{Ren}},
  \bibinfo{author}{\bibfnamefont{Y.}~\bibnamefont{Li}},
  \bibinfo{author}{\bibfnamefont{Y.}~\bibnamefont{Wang}},
  \bibinfo{author}{\bibfnamefont{X.}~\bibnamefont{Lin}},
  \bibinfo{author}{\bibfnamefont{Y.}~\bibnamefont{Luo}},
  \bibinfo{author}{\bibfnamefont{S.}~\bibnamefont{Jiang}},
  \bibnamefont{et~al.}, \bibinfo{journal}{EPL} \textbf{\bibinfo{volume}{83}},
  \bibinfo{pages}{67006} (\bibinfo{year}{2008}).

\bibitem[{\citenamefont{Hsu et~al.}(2008)\citenamefont{Hsu, Luo, Yeh, Chen,
  Huang, Wu, Lee, Huang, Chu, Yan et~al.}}]{hsu2008superconductivity}
\bibinfo{author}{\bibfnamefont{F.-C.} \bibnamefont{Hsu}},
  \bibinfo{author}{\bibfnamefont{J.-Y.} \bibnamefont{Luo}},
  \bibinfo{author}{\bibfnamefont{K.-W.} \bibnamefont{Yeh}},
  \bibinfo{author}{\bibfnamefont{T.-K.} \bibnamefont{Chen}},
  \bibinfo{author}{\bibfnamefont{T.-W.} \bibnamefont{Huang}},
  \bibinfo{author}{\bibfnamefont{P.~M.} \bibnamefont{Wu}},
  \bibinfo{author}{\bibfnamefont{Y.-C.} \bibnamefont{Lee}},
  \bibinfo{author}{\bibfnamefont{Y.-L.} \bibnamefont{Huang}},
  \bibinfo{author}{\bibfnamefont{Y.-Y.} \bibnamefont{Chu}},
  \bibinfo{author}{\bibfnamefont{D.-C.} \bibnamefont{Yan}},
  \bibnamefont{et~al.}, \bibinfo{journal}{Proceedings of the National Academy
  of Sciences} \textbf{\bibinfo{volume}{105}}, \bibinfo{pages}{14262}
  (\bibinfo{year}{2008}).

\bibitem[{\citenamefont{Song et~al.}(2011)\citenamefont{Song, Wang, Jiang, Li,
  Wang, He, Chen, Ma, and Xue}}]{PhysRevB.84.020503}
\bibinfo{author}{\bibfnamefont{C.-L.} \bibnamefont{Song}},
  \bibinfo{author}{\bibfnamefont{Y.-L.} \bibnamefont{Wang}},
  \bibinfo{author}{\bibfnamefont{Y.-P.} \bibnamefont{Jiang}},
  \bibinfo{author}{\bibfnamefont{Z.}~\bibnamefont{Li}},
  \bibinfo{author}{\bibfnamefont{L.}~\bibnamefont{Wang}},
  \bibinfo{author}{\bibfnamefont{K.}~\bibnamefont{He}},
  \bibinfo{author}{\bibfnamefont{X.}~\bibnamefont{Chen}},
  \bibinfo{author}{\bibfnamefont{X.-C.} \bibnamefont{Ma}}, \bibnamefont{and}
  \bibinfo{author}{\bibfnamefont{Q.-K.} \bibnamefont{Xue}},
  \bibinfo{journal}{Phys. Rev. B} \textbf{\bibinfo{volume}{84}},
  \bibinfo{pages}{020503} (\bibinfo{year}{2011}).

\bibitem[{\citenamefont{Lee et~al.}(2014)\citenamefont{Lee, Schmitt, Moore,
  Johnston, Cui, Li, Yi, Liu, Hashimoto, Zhang et~al.}}]{lee2014interfacial}
\bibinfo{author}{\bibfnamefont{J.}~\bibnamefont{Lee}},
  \bibinfo{author}{\bibfnamefont{F.}~\bibnamefont{Schmitt}},
  \bibinfo{author}{\bibfnamefont{R.}~\bibnamefont{Moore}},
  \bibinfo{author}{\bibfnamefont{S.}~\bibnamefont{Johnston}},
  \bibinfo{author}{\bibfnamefont{Y.-T.} \bibnamefont{Cui}},
  \bibinfo{author}{\bibfnamefont{W.}~\bibnamefont{Li}},
  \bibinfo{author}{\bibfnamefont{M.}~\bibnamefont{Yi}},
  \bibinfo{author}{\bibfnamefont{Z.}~\bibnamefont{Liu}},
  \bibinfo{author}{\bibfnamefont{M.}~\bibnamefont{Hashimoto}},
  \bibinfo{author}{\bibfnamefont{Y.}~\bibnamefont{Zhang}},
  \bibnamefont{et~al.}, \bibinfo{journal}{Nature}
  \textbf{\bibinfo{volume}{515}}, \bibinfo{pages}{245} (\bibinfo{year}{2014}).

\bibitem[{\citenamefont{Peng et~al.}(2014)\citenamefont{Peng, Shen, Xie, Xu,
  Tan, Xia, Zhang, Cao, Gong, Hu et~al.}}]{PhysRevLett.112.107001}
\bibinfo{author}{\bibfnamefont{R.}~\bibnamefont{Peng}},
  \bibinfo{author}{\bibfnamefont{X.~P.} \bibnamefont{Shen}},
  \bibinfo{author}{\bibfnamefont{X.}~\bibnamefont{Xie}},
  \bibinfo{author}{\bibfnamefont{H.~C.} \bibnamefont{Xu}},
  \bibinfo{author}{\bibfnamefont{S.~Y.} \bibnamefont{Tan}},
  \bibinfo{author}{\bibfnamefont{M.}~\bibnamefont{Xia}},
  \bibinfo{author}{\bibfnamefont{T.}~\bibnamefont{Zhang}},
  \bibinfo{author}{\bibfnamefont{H.~Y.} \bibnamefont{Cao}},
  \bibinfo{author}{\bibfnamefont{X.~G.} \bibnamefont{Gong}},
  \bibinfo{author}{\bibfnamefont{J.~P.} \bibnamefont{Hu}},
  \bibnamefont{et~al.}, \bibinfo{journal}{Phys. Rev. Lett.}
  \textbf{\bibinfo{volume}{112}}, \bibinfo{pages}{107001}
  (\bibinfo{year}{2014}).

\bibitem[{\citenamefont{Dai et~al.}(2012)\citenamefont{Dai, Hu, and
  Dagotto}}]{dai2012magnetism}
\bibinfo{author}{\bibfnamefont{P.}~\bibnamefont{Dai}},
  \bibinfo{author}{\bibfnamefont{J.}~\bibnamefont{Hu}}, \bibnamefont{and}
  \bibinfo{author}{\bibfnamefont{E.}~\bibnamefont{Dagotto}},
  \bibinfo{journal}{Nature Physics} \textbf{\bibinfo{volume}{8}},
  \bibinfo{pages}{709} (\bibinfo{year}{2012}).

\bibitem[{\citenamefont{de~La~Cruz et~al.}(2008)\citenamefont{de~La~Cruz,
  Huang, Lynn, Li, Ratcliff~II, Zarestky, Mook, Chen, Luo, Wang
  et~al.}}]{de2008magnetic}
\bibinfo{author}{\bibfnamefont{C.}~\bibnamefont{de~La~Cruz}},
  \bibinfo{author}{\bibfnamefont{Q.}~\bibnamefont{Huang}},
  \bibinfo{author}{\bibfnamefont{J.}~\bibnamefont{Lynn}},
  \bibinfo{author}{\bibfnamefont{J.}~\bibnamefont{Li}},
  \bibinfo{author}{\bibfnamefont{W.}~\bibnamefont{Ratcliff~II}},
  \bibinfo{author}{\bibfnamefont{J.~L.} \bibnamefont{Zarestky}},
  \bibinfo{author}{\bibfnamefont{H.}~\bibnamefont{Mook}},
  \bibinfo{author}{\bibfnamefont{G.}~\bibnamefont{Chen}},
  \bibinfo{author}{\bibfnamefont{J.}~\bibnamefont{Luo}},
  \bibinfo{author}{\bibfnamefont{N.}~\bibnamefont{Wang}}, \bibnamefont{et~al.},
  \bibinfo{journal}{Nature(London)} \textbf{\bibinfo{volume}{453}},
  \bibinfo{pages}{899} (\bibinfo{year}{2008}).

\bibitem[{\citenamefont{Zhao et~al.}(2008)\citenamefont{Zhao, Huang,
  de~La~Cruz, Li, Lynn, Chen, Green, Chen, Li, Li et~al.}}]{zhao2008structural}
\bibinfo{author}{\bibfnamefont{J.}~\bibnamefont{Zhao}},
  \bibinfo{author}{\bibfnamefont{Q.}~\bibnamefont{Huang}},
  \bibinfo{author}{\bibfnamefont{C.}~\bibnamefont{de~La~Cruz}},
  \bibinfo{author}{\bibfnamefont{S.}~\bibnamefont{Li}},
  \bibinfo{author}{\bibfnamefont{J.}~\bibnamefont{Lynn}},
  \bibinfo{author}{\bibfnamefont{Y.}~\bibnamefont{Chen}},
  \bibinfo{author}{\bibfnamefont{M.}~\bibnamefont{Green}},
  \bibinfo{author}{\bibfnamefont{G.}~\bibnamefont{Chen}},
  \bibinfo{author}{\bibfnamefont{G.}~\bibnamefont{Li}},
  \bibinfo{author}{\bibfnamefont{Z.}~\bibnamefont{Li}}, \bibnamefont{et~al.},
  \bibinfo{journal}{Nature materials} \textbf{\bibinfo{volume}{7}},
  \bibinfo{pages}{953} (\bibinfo{year}{2008}).

\bibitem[{\citenamefont{Dong et~al.}(2008)\citenamefont{Dong, Zhang, Xu, Li,
  Li, Hu, Wu, Chen, Dai, Luo et~al.}}]{0295-5075-83-2-27006}
\bibinfo{author}{\bibfnamefont{J.}~\bibnamefont{Dong}},
  \bibinfo{author}{\bibfnamefont{H.~J.} \bibnamefont{Zhang}},
  \bibinfo{author}{\bibfnamefont{G.}~\bibnamefont{Xu}},
  \bibinfo{author}{\bibfnamefont{Z.}~\bibnamefont{Li}},
  \bibinfo{author}{\bibfnamefont{G.}~\bibnamefont{Li}},
  \bibinfo{author}{\bibfnamefont{W.~Z.} \bibnamefont{Hu}},
  \bibinfo{author}{\bibfnamefont{D.}~\bibnamefont{Wu}},
  \bibinfo{author}{\bibfnamefont{G.~F.} \bibnamefont{Chen}},
  \bibinfo{author}{\bibfnamefont{X.}~\bibnamefont{Dai}},
  \bibinfo{author}{\bibfnamefont{J.~L.} \bibnamefont{Luo}},
  \bibnamefont{et~al.}, \bibinfo{journal}{EPL} \textbf{\bibinfo{volume}{83}},
  \bibinfo{pages}{27006} (\bibinfo{year}{2008}).

\bibitem[{\citenamefont{Uemura}(2009)}]{uemura2009superconductivity}
\bibinfo{author}{\bibfnamefont{Y.~J.} \bibnamefont{Uemura}},
  \bibinfo{journal}{Nature materials} \textbf{\bibinfo{volume}{8}},
  \bibinfo{pages}{253} (\bibinfo{year}{2009}).

\bibitem[{\citenamefont{Ma and Lu}(2008)}]{ma2008iron}
\bibinfo{author}{\bibfnamefont{F.}~\bibnamefont{Ma}} \bibnamefont{and}
  \bibinfo{author}{\bibfnamefont{Z.-Y.} \bibnamefont{Lu}},
  \bibinfo{journal}{Phys. Rev. B} \textbf{\bibinfo{volume}{78}},
  \bibinfo{pages}{033111} (\bibinfo{year}{2008}).

\bibitem[{\citenamefont{Yildirim}(2008)}]{yildirim2008origin}
\bibinfo{author}{\bibfnamefont{T.}~\bibnamefont{Yildirim}},
  \bibinfo{journal}{Physical Review Letters} \textbf{\bibinfo{volume}{101}},
  \bibinfo{pages}{057010} (\bibinfo{year}{2008}).

\bibitem[{\citenamefont{Cao et~al.}(2014)\citenamefont{Cao, Tan, Xiang, Feng,
  and Gong}}]{cao2014interfacial}
\bibinfo{author}{\bibfnamefont{H.-Y.} \bibnamefont{Cao}},
  \bibinfo{author}{\bibfnamefont{S.}~\bibnamefont{Tan}},
  \bibinfo{author}{\bibfnamefont{H.}~\bibnamefont{Xiang}},
  \bibinfo{author}{\bibfnamefont{D.}~\bibnamefont{Feng}}, \bibnamefont{and}
  \bibinfo{author}{\bibfnamefont{X.-G.} \bibnamefont{Gong}},
  \bibinfo{journal}{Physical Review B} \textbf{\bibinfo{volume}{89}},
  \bibinfo{pages}{014501} (\bibinfo{year}{2014}).

\bibitem[{\citenamefont{Glasbrenner et~al.}(2015)\citenamefont{Glasbrenner,
  Mazin, Jeschke, Hirschfeld, Fernandes, and
  Valent{\'\i}}}]{glasbrenner2015effect}
\bibinfo{author}{\bibfnamefont{J.}~\bibnamefont{Glasbrenner}},
  \bibinfo{author}{\bibfnamefont{I.}~\bibnamefont{Mazin}},
  \bibinfo{author}{\bibfnamefont{H.~O.} \bibnamefont{Jeschke}},
  \bibinfo{author}{\bibfnamefont{P.}~\bibnamefont{Hirschfeld}},
  \bibinfo{author}{\bibfnamefont{R.}~\bibnamefont{Fernandes}},
  \bibnamefont{and}
  \bibinfo{author}{\bibfnamefont{R.}~\bibnamefont{Valent{\'\i}}},
  \bibinfo{journal}{Nature Physics}  (\bibinfo{year}{2015}).

\bibitem[{\citenamefont{Cao et~al.}(2015)\citenamefont{Cao, Chen, Xiang, and
  Gong}}]{PhysRevB.91.020504}
\bibinfo{author}{\bibfnamefont{H.-Y.} \bibnamefont{Cao}},
  \bibinfo{author}{\bibfnamefont{S.}~\bibnamefont{Chen}},
  \bibinfo{author}{\bibfnamefont{H.}~\bibnamefont{Xiang}}, \bibnamefont{and}
  \bibinfo{author}{\bibfnamefont{X.-G.} \bibnamefont{Gong}},
  \bibinfo{journal}{Phys. Rev. B} \textbf{\bibinfo{volume}{91}},
  \bibinfo{pages}{020504} (\bibinfo{year}{2015}).

\bibitem[{\citenamefont{McQueen et~al.}(2009)\citenamefont{McQueen, Williams,
  Stephens, Tao, Zhu, Ksenofontov, Casper, Felser, and
  Cava}}]{PhysRevLett.103.057002}
\bibinfo{author}{\bibfnamefont{T.~M.} \bibnamefont{McQueen}},
  \bibinfo{author}{\bibfnamefont{A.~J.} \bibnamefont{Williams}},
  \bibinfo{author}{\bibfnamefont{P.~W.} \bibnamefont{Stephens}},
  \bibinfo{author}{\bibfnamefont{J.}~\bibnamefont{Tao}},
  \bibinfo{author}{\bibfnamefont{Y.}~\bibnamefont{Zhu}},
  \bibinfo{author}{\bibfnamefont{V.}~\bibnamefont{Ksenofontov}},
  \bibinfo{author}{\bibfnamefont{F.}~\bibnamefont{Casper}},
  \bibinfo{author}{\bibfnamefont{C.}~\bibnamefont{Felser}}, \bibnamefont{and}
  \bibinfo{author}{\bibfnamefont{R.~J.} \bibnamefont{Cava}},
  \bibinfo{journal}{Phys. Rev. Lett.} \textbf{\bibinfo{volume}{103}},
  \bibinfo{pages}{057002} (\bibinfo{year}{2009}).

\bibitem[{\citenamefont{Yaresko et~al.}(2009)\citenamefont{Yaresko, Liu,
  Antonov, and Andersen}}]{YareskoInterplay}
\bibinfo{author}{\bibfnamefont{A.~N.} \bibnamefont{Yaresko}},
  \bibinfo{author}{\bibfnamefont{G.-Q.} \bibnamefont{Liu}},
  \bibinfo{author}{\bibfnamefont{V.~N.} \bibnamefont{Antonov}},
  \bibnamefont{and} \bibinfo{author}{\bibfnamefont{O.~K.}
  \bibnamefont{Andersen}}, \bibinfo{journal}{Phys. Rev. B}
  \textbf{\bibinfo{volume}{79}}, \bibinfo{pages}{144421}
  (\bibinfo{year}{2009}).

\bibitem[{\citenamefont{Wysocki et~al.}(2011)\citenamefont{Wysocki,
  Belashchenko, and Antropov}}]{wysocki2011consistent}
\bibinfo{author}{\bibfnamefont{A.~L.} \bibnamefont{Wysocki}},
  \bibinfo{author}{\bibfnamefont{K.~D.} \bibnamefont{Belashchenko}},
  \bibnamefont{and} \bibinfo{author}{\bibfnamefont{V.~P.}
  \bibnamefont{Antropov}}, \bibinfo{journal}{Nature Physics}
  \textbf{\bibinfo{volume}{7}}, \bibinfo{pages}{485} (\bibinfo{year}{2011}).

\bibitem[{\citenamefont{Hobbs et~al.}(2000)\citenamefont{Hobbs, Kresse, and
  Hafner}}]{PhysRevB.62.11556}
\bibinfo{author}{\bibfnamefont{D.}~\bibnamefont{Hobbs}},
  \bibinfo{author}{\bibfnamefont{G.}~\bibnamefont{Kresse}}, \bibnamefont{and}
  \bibinfo{author}{\bibfnamefont{J.}~\bibnamefont{Hafner}},
  \bibinfo{journal}{Phys. Rev. B} \textbf{\bibinfo{volume}{62}},
  \bibinfo{pages}{11556} (\bibinfo{year}{2000}).

\bibitem[{\citenamefont{Kresse and Furthmüller}(1996)}]{Kresse199615}
\bibinfo{author}{\bibfnamefont{G.}~\bibnamefont{Kresse}} \bibnamefont{and}
  \bibinfo{author}{\bibfnamefont{J.}~\bibnamefont{Furthmüller}},
  \bibinfo{journal}{Computational Materials Science}
  \textbf{\bibinfo{volume}{6}}, \bibinfo{pages}{15 } (\bibinfo{year}{1996}),
  ISSN \bibinfo{issn}{0927-0256}.

\bibitem[{\citenamefont{Kresse and Furthm\"uller}(1996)}]{PhysRevB.54.11169}
\bibinfo{author}{\bibfnamefont{G.}~\bibnamefont{Kresse}} \bibnamefont{and}
  \bibinfo{author}{\bibfnamefont{J.}~\bibnamefont{Furthm\"uller}},
  \bibinfo{journal}{Phys. Rev. B} \textbf{\bibinfo{volume}{54}},
  \bibinfo{pages}{11169} (\bibinfo{year}{1996}).

\bibitem[{\citenamefont{Bl\"ochl}(1994)}]{PhysRevB.50.17953}
\bibinfo{author}{\bibfnamefont{P.~E.} \bibnamefont{Bl\"ochl}},
  \bibinfo{journal}{Phys. Rev. B} \textbf{\bibinfo{volume}{50}},
  \bibinfo{pages}{17953} (\bibinfo{year}{1994}).

\bibitem[{\citenamefont{Kresse and Joubert}(1999)}]{PhysRevB.59.1758}
\bibinfo{author}{\bibfnamefont{G.}~\bibnamefont{Kresse}} \bibnamefont{and}
  \bibinfo{author}{\bibfnamefont{D.}~\bibnamefont{Joubert}},
  \bibinfo{journal}{Phys. Rev. B} \textbf{\bibinfo{volume}{59}},
  \bibinfo{pages}{1758} (\bibinfo{year}{1999}).

\bibitem[{\citenamefont{Perdew et~al.}(1996)\citenamefont{Perdew, Burke, and
  Ernzerhof}}]{PhysRevLett.77.3865}
\bibinfo{author}{\bibfnamefont{J.~P.} \bibnamefont{Perdew}},
  \bibinfo{author}{\bibfnamefont{K.}~\bibnamefont{Burke}}, \bibnamefont{and}
  \bibinfo{author}{\bibfnamefont{M.}~\bibnamefont{Ernzerhof}},
  \bibinfo{journal}{Phys. Rev. Lett.} \textbf{\bibinfo{volume}{77}},
  \bibinfo{pages}{3865} (\bibinfo{year}{1996}).

\bibitem[{\citenamefont{Monkhorst and Pack}(1976)}]{PhysRevB.13.5188}
\bibinfo{author}{\bibfnamefont{H.~J.} \bibnamefont{Monkhorst}}
  \bibnamefont{and} \bibinfo{author}{\bibfnamefont{J.~D.} \bibnamefont{Pack}},
  \bibinfo{journal}{Phys. Rev. B} \textbf{\bibinfo{volume}{13}},
  \bibinfo{pages}{5188} (\bibinfo{year}{1976}).

\bibitem[{\citenamefont{Si and Abrahams}(2008)}]{si2008strong}
\bibinfo{author}{\bibfnamefont{Q.}~\bibnamefont{Si}} \bibnamefont{and}
  \bibinfo{author}{\bibfnamefont{E.}~\bibnamefont{Abrahams}},
  \bibinfo{journal}{Physical Review Letters} \textbf{\bibinfo{volume}{101}},
  \bibinfo{pages}{076401} (\bibinfo{year}{2008}).

\bibitem[{\citenamefont{Yao and Carlson}(2008)}]{yaomagnetic}
\bibinfo{author}{\bibfnamefont{D.-X.} \bibnamefont{Yao}} \bibnamefont{and}
  \bibinfo{author}{\bibfnamefont{E.~W.} \bibnamefont{Carlson}},
  \bibinfo{journal}{Phys. Rev. B} \textbf{\bibinfo{volume}{78}},
  \bibinfo{pages}{052507} (\bibinfo{year}{2008}).

\bibitem[{\citenamefont{Fang et~al.}(2008)\citenamefont{Fang, Yao, Tsai, Hu,
  and Kivelson}}]{fangtheory}
\bibinfo{author}{\bibfnamefont{C.}~\bibnamefont{Fang}},
  \bibinfo{author}{\bibfnamefont{H.}~\bibnamefont{Yao}},
  \bibinfo{author}{\bibfnamefont{W.-F.} \bibnamefont{Tsai}},
  \bibinfo{author}{\bibfnamefont{J.}~\bibnamefont{Hu}}, \bibnamefont{and}
  \bibinfo{author}{\bibfnamefont{S.~A.} \bibnamefont{Kivelson}},
  \bibinfo{journal}{Phys. Rev. B} \textbf{\bibinfo{volume}{77}},
  \bibinfo{pages}{224509} (\bibinfo{year}{2008}).

\bibitem[{\citenamefont{Ma et~al.}(2008)\citenamefont{Ma, Lu, and
  Xiang}}]{ma2008arsenic}
\bibinfo{author}{\bibfnamefont{F.}~\bibnamefont{Ma}},
  \bibinfo{author}{\bibfnamefont{Z.-Y.} \bibnamefont{Lu}}, \bibnamefont{and}
  \bibinfo{author}{\bibfnamefont{T.}~\bibnamefont{Xiang}},
  \bibinfo{journal}{Physical Review B} \textbf{\bibinfo{volume}{78}},
  \bibinfo{pages}{224517} (\bibinfo{year}{2008}).

\bibitem[{\citenamefont{Jiang et~al.}(2009)\citenamefont{Jiang, Kr\"uger,
  Moore, Sheng, Zaanen, and Weng}}]{PhysRevB.79.174409}
\bibinfo{author}{\bibfnamefont{H.~C.} \bibnamefont{Jiang}},
  \bibinfo{author}{\bibfnamefont{F.}~\bibnamefont{Kr\"uger}},
  \bibinfo{author}{\bibfnamefont{J.~E.} \bibnamefont{Moore}},
  \bibinfo{author}{\bibfnamefont{D.~N.} \bibnamefont{Sheng}},
  \bibinfo{author}{\bibfnamefont{J.}~\bibnamefont{Zaanen}}, \bibnamefont{and}
  \bibinfo{author}{\bibfnamefont{Z.~Y.} \bibnamefont{Weng}},
  \bibinfo{journal}{Phys. Rev. B} \textbf{\bibinfo{volume}{79}},
  \bibinfo{pages}{174409} (\bibinfo{year}{2009}).

\bibitem[{\citenamefont{Wang et~al.}(2015)\citenamefont{Wang, Kivelson, and
  Lee}}]{wang2015nematicity}
\bibinfo{author}{\bibfnamefont{F.}~\bibnamefont{Wang}},
  \bibinfo{author}{\bibfnamefont{S.~A.} \bibnamefont{Kivelson}},
  \bibnamefont{and} \bibinfo{author}{\bibfnamefont{D.-H.} \bibnamefont{Lee}},
  \bibinfo{journal}{Nature Physics}  (\bibinfo{year}{2015}).

\bibitem[{\citenamefont{Yu and Si}(2015)}]{yuantiferroquadrupolar}
\bibinfo{author}{\bibfnamefont{R.}~\bibnamefont{Yu}} \bibnamefont{and}
  \bibinfo{author}{\bibfnamefont{Q.}~\bibnamefont{Si}}, \bibinfo{journal}{Phys.
  Rev. Lett.} \textbf{\bibinfo{volume}{115}}, \bibinfo{pages}{116401}
  (\bibinfo{year}{2015}).

\bibitem[{\citenamefont{Chubukov et~al.}(2015)\citenamefont{Chubukov,
  Fernandes, and Schmalian}}]{PhysRevB.91.201105}
\bibinfo{author}{\bibfnamefont{A.~V.} \bibnamefont{Chubukov}},
  \bibinfo{author}{\bibfnamefont{R.~M.} \bibnamefont{Fernandes}},
  \bibnamefont{and}
  \bibinfo{author}{\bibfnamefont{J.}~\bibnamefont{Schmalian}},
  \bibinfo{journal}{Phys. Rev. B} \textbf{\bibinfo{volume}{91}},
  \bibinfo{pages}{201105} (\bibinfo{year}{2015}).

\bibitem[{Xpr(2015)}]{Xproceedings}
\emph{\bibinfo{title}{Presentation by Zhong-Yi Lu at Chinese Physics Society
  Meeting}} (\bibinfo{year}{2015}).

\bibitem[{\citenamefont{Huang et~al.}(2008)\citenamefont{Huang, Qiu, Bao,
  Green, Lynn, Gasparovic, Wu, Wu, and Chen}}]{huang2008neutron}
\bibinfo{author}{\bibfnamefont{Q.}~\bibnamefont{Huang}},
  \bibinfo{author}{\bibfnamefont{Y.}~\bibnamefont{Qiu}},
  \bibinfo{author}{\bibfnamefont{W.}~\bibnamefont{Bao}},
  \bibinfo{author}{\bibfnamefont{M.}~\bibnamefont{Green}},
  \bibinfo{author}{\bibfnamefont{J.}~\bibnamefont{Lynn}},
  \bibinfo{author}{\bibfnamefont{Y.}~\bibnamefont{Gasparovic}},
  \bibinfo{author}{\bibfnamefont{T.}~\bibnamefont{Wu}},
  \bibinfo{author}{\bibfnamefont{G.}~\bibnamefont{Wu}}, \bibnamefont{and}
  \bibinfo{author}{\bibfnamefont{X.}~\bibnamefont{Chen}},
  \bibinfo{journal}{Physical Review Letters} \textbf{\bibinfo{volume}{101}},
  \bibinfo{pages}{257003} (\bibinfo{year}{2008}).

\bibitem[{\citenamefont{Ma et~al.}(2009)\citenamefont{Ma, Ji, Hu, Lu, and
  Xiang}}]{PhysRevLett.102.177003}
\bibinfo{author}{\bibfnamefont{F.}~\bibnamefont{Ma}},
  \bibinfo{author}{\bibfnamefont{W.}~\bibnamefont{Ji}},
  \bibinfo{author}{\bibfnamefont{J.}~\bibnamefont{Hu}},
  \bibinfo{author}{\bibfnamefont{Z.-Y.} \bibnamefont{Lu}}, \bibnamefont{and}
  \bibinfo{author}{\bibfnamefont{T.}~\bibnamefont{Xiang}},
  \bibinfo{journal}{Phys. Rev. Lett.} \textbf{\bibinfo{volume}{102}},
  \bibinfo{pages}{177003} (\bibinfo{year}{2009}).

\end{thebibliography}
\end{document}